\documentclass[superscriptaddress,aps,preprintnumbers,amsmath,showpacs,amssymb,nofootinbib,reprint]{revtex4-1}
\usepackage{bm, color} 
\usepackage{amssymb,amsfonts,slashed,amsthm,amsmath,graphicx, soul}
\usepackage{subfigure}
\usepackage{hyperref}
\usepackage{tikz,xcolor}

\begin{document}

\title{Emission of Nambu-Goldstone bosons from the semilocal string network}

\author{Yukihiro Kanda}
\affiliation{Institute for Cosmic Ray Research, The University of Tokyo, \\
Kashiwa, Chiba, 277-8582, Japan}

\author{Naoya Kitajima}
\affiliation{Frontier Research Institute for Interdisciplinary Sciences, Tohoku University, \\
6-3 Azaaoba, Aramaki, Aoba-ku, Sendai 980-8578, Japan}
\affiliation{Department of Physics, Tohoku University, \\
6-3 Azaaoba, Aramaki, Aoba-ku, Sendai 980-8578, Japan}

\preprint{TU-1282}

\begin{abstract}
Semilocal cosmic string is a line-like non-topological soliton associated with the breakdown of the $SU(2)_{\rm global} \times U(1)_{\rm gauge}$ symmetry to the $U(1)_{\rm global}$ symmetry. The broken phase has two massless Nambu-Goldstone (NG) modes as dynamical fields, and they can be emitted by semilocal strings.
In this paper, we numerically show that such NG bosons are copiously produced with the evolution of the semilocal string network in the early universe.
Our numerical analysis shows that the spectrum of produced particles has a peak at low momenta corresponding to the horizon scale. 
If the emitted NG bosons acquire mass due to soft-breaking terms, they can take the role of dark matter. This scenario typically predicts very light pseudo NG boson dark matter.
\end{abstract}

\maketitle

\section{Introduction}
The cosmic string is a line-like topological defect associated with the breakdown of continuous U(1) symmetry \cite{Kibble:1976sj,Vilenkin:2000jqa}. The early time evolution of the Universe allows the spontaneous breaking of such a symmetry and then cosmic string forms a network. The network evolves following the so-called scaling regime \cite{Kibble:1984hp,Bennett:1985qt,Bennett:1986zn,Bennett:1989yp}. A remarkable aspect of the cosmic sting is the emission of gravitational waves \cite{Vilenkin:1981bx,Vachaspati:1984gt}. In particular, the cosmic string associated with the U(1) gauge symmetry breaking predicts the nearly scale-invariant spectrum of gravitational waves, and it is one of the main targets for gravitational wave observations. See \cite{Cui:2017ufi,Caprini:2018mtu,Auclair:2019wcv} for reviews.

Cosmic strings can exist even if their stability is not ensured by the topology of the vacuum manifold. The semilocal string is one of the representative and simplest examples \cite{Vachaspati:1991dz,Hindmarsh:1991jq,Hindmarsh:1992yy,Preskill:1992bf,Gibbons:1992gt,Achucarro:1998ux,Achucarro:1999it}. The semilocal string arises when the $SU(2)_{\rm global} \times U(1)_{\rm gauge}$ symmetry is broken to the residual $U(1)_{\rm global}$ symmetry.
The topology of the vacuum manifold in this model is $S_3$, which cannot stabilize the string, but it has been shown that the semilocal string is stable when the gauge field is heavier than the scalar radial mode \cite{Hindmarsh:1991jq}. 

Furthermore, unlike the Abelian-Higgs model, this model contains massless degrees of freedom associated with Nambu-Goldstone (NG) bosons after the symmetry breaking.
Thus, the semilocal string has a channel to decay into those light particles. This is similar to the emission of the (QCD) axions from global (axion) strings, which is extensively studied in the literature \cite{Davis:1986xc,Vilenkin:1986ku,Garfinkle:1987yw,Yamaguchi:1998gx,Hagmann:2000ja,Hiramatsu:2010yu,Fleury:2015aca,Kawasaki:2018bzv,Gorghetto:2018myk,Gorghetto:2020qws,Buschmann:2021sdq,Saikawa:2024bta,Kim:2024wku,Kim:2024dtq,Benabou:2024msj}. In this scenario, the axion can account for the dark matter as it acquires the mass in the subsequent cosmological evolutions through the non-perturvative effect. 
Since the topological defects emit particles with low momenta, which is typically the Hubble scale at the emission, the produced particles become non-relativistic and thus behave as cold dark matter. This non-thermal production allows for the dark matter to be very light in contrast with the thermal production.

Similarly, the emission of light gauge bosons is also kinematically allowed in the case with the Abelian-Higgs string if the gauge coupling constant is much smaller than the scalar self-coupling constant which corresponds to the type-II string \cite{Long:2019lwl,Kitajima:2022lre,Kitajima:2023vre}. In this case, the dark photon can account for the light dark matter. In addition, the Abelian-Higgs string network can also produce axions through the axion-gauge coupling \cite{Kitajima:2025jct}.

The light (massless) particle emission is essential in the context of the gravitational wave emissivity from cosmic strings. 
Notably, if the system allows such light particle emissions, the decay channel to the gravitational wave is significantly suppressed, which drastically changes the resultant gravitational wave spectrum \cite{Chang:2019mza,Chang:2021afa,Gorghetto:2021fsn,Kitajima:2022lre}.
Thus, it is important to investigate the efficiency of the light (massless) particle emission from the string network quantitatively.

In this paper, we show by numerical simulations that the massless NG bosons can be copiously produced from the network of semilocal strings.
Our simulations exhibit that the produced NG bosons have low momenta corresponding to the Hubble scale at their productions, similar to the case with axion/dark photon emissions mentioned above. In addition, we show that those NG bosons can explain the relic dark matter abundance if they acquire the mass due to soft-breaking terms.

We present the model in Sec.\ref{sec:model} and show the decomposition of the field contents into massless NG modes and a massive radial mode in Sec.\ref{sec:NGmode}. Sec.\ref{sec:numerical} contains the setup and results of our numerical simulations. Finally, we give discussions in Sec.\ref{sec:discussion}.

\section{Model} \label{sec:model}
Let us consider the model based on the $SU(2)_{\rm global} \times U(1)_{\rm gauge}$ symmetry\cite{Vachaspati:1991dz,Achucarro:1999it}, described by the following Lagrangian,
\begin{align}
\mathcal{L} = D_\mu \Phi^\dagger D^\mu \Phi+V(\Phi)+\frac{1}{4}F_{\mu\nu}F^{\mu\nu},
\end{align}
where $\Phi = (\Phi_1,\Phi_2)^T$ is a $SU(2)$ doublet complex scalar field, $D_\mu = \partial_\mu -iqA_\mu$ is the covariant derivative associated with the U(1) gauge field, $A_\mu$, with $q$ being the charge of the scalar field and $F_{\mu\nu} = \partial_\mu A_\nu - \partial_\nu A_\mu$ is the field strength tensor.
We take the implicit sum over the repeated indices.
The potential for the scalar field is given by
\begin{align}
V(\Phi) = \lambda \left(\Phi^\dagger \Phi-\frac{v^2}{2} \right)^2,
\end{align}
where $\lambda$ is a dimensionless coupling constant and $v$ is the vacuum expectation value (VEV) of the scalar field.

The vacuum manifold in this model is $S^3$, and thus its fundamental group is trivial. However, it is known that the following string solution exists:
\begin{align}
\label{semilocal_string_solution}
    \Phi =  \frac{v}{\sqrt{2}} f(r) e^{iw\theta} n, \quad A_\mu = \delta_{\mu\theta} \frac{wg(r)}{qr},
\end{align}
where $n$ is a normalized doublet satisfying $n^\dagger n=1$, $w$ is a winding number, and we use cylindrical coordinates $(t,r,\theta,z)$. The functions $f(r)$ and $g(r)$ are monotonically increasing, satisfying $f(0)=g(0)=0$ and $f(\infty)=g(\infty)=1$. This solution is known as the semilocal string\cite{Vachaspati:1991dz} and it is stable when $\beta=2\lambda/q^2<1$~\cite{Hindmarsh:1991jq}.

Assuming the Friedmann-Lema\^{i}tre-Robertson-Walker (FLRW) Universe with zero spatial curvature, one can derive the following system of evolution equations and the constraint equation for the scalar and gauge fields,
\begin{align}
&\ddot{\Phi}_c + 3H \dot{\Phi}_c - \frac{D_i D_i \Phi_c}{a^2} + \frac{\partial V}{\partial \Phi^*_c} = 0, \label{eq:EoM_Phi}\\
&\ddot{A}_i+H\dot{A}_i-\frac{1}{a^2}(\nabla^2 A_i -\partial_i \partial_j A_j) -2q {\rm Im} (\Phi_c^* D_i \Phi_c) = 0, \label{eq:EoM_A}\\
&\partial_i \dot{A}_i - 2qa^2 {\rm Im}(\Phi_c^* \dot{\Phi}_c) = 0 \label{eq:Gauss_law}
\end{align}
where $c=1,2$, $H = \dot{a}/a$ is the Hubble parameter, $a$ is the scale factor and the overdot represents the derivative with respect to the cosmic time, $t$. Here and in what follows, we impose the temporal gauge condition, $A_0 = 0$, and assume the radiation-dominated universe, which reads $H = 1/(2t)$ and $a \propto t^{1/2}$.

\section{Massless NG mode} \label{sec:NGmode}
In this model, there are two massless NG modes associated with the global SU(2) symmetry breaking.
First, let us decompose two individual scalar fields into the radial component, $\varphi_c$, and the phase component, $\theta_c$, as follows,
\begin{align}
\Phi_c = \frac{1}{\sqrt{2}} \varphi_c e^{i\theta_c}.
\end{align}
Then, let us define the following linear combinations of the phase components for complex scalar fields,
\begin{align}
\theta_\pm = \frac{\theta_1 \pm \theta_2}{2}.
\end{align}
Note that the first (plus) mode corresponds to the $U(1)$ gauge degree of freedom and contributes to the longitudinal mode of the massive gauge field after the symmetry breaking.
On the other hand, the second (minus) mode corresponds to $U(1)$ global symmetry associated with the relative phase between $\Phi_1$ and $\Phi_2$ and thus remains massless after the symmetry breaking.

The second massless mode corresponds to the degree of freedom of changing the combination of $(\varphi_1,\varphi_2)$ without changing the norm $\varphi_1^2 + \varphi_2^2$, and thus, this mode remains on the vacuum manifold.
Note that the mode changing the norm is clearly a massive mode because the potential $V \propto ( \varphi_1^2 + \varphi_2^2 - v^2)^2$ depends only on this mode.
To extract the second massless mode, let us consider the $\varphi_1$-$\varphi_2$ plane and introduce the radial coordinate, $\varphi_r$, and the angular coordinate, $\vartheta$, on this field plane.
Namely, they are defined by
\begin{align}
\varphi_1 = \varphi_r \cos\vartheta,~~~ \varphi_2 = \varphi_r \sin \vartheta.
\end{align}
Clearly, $\varphi_r = \sqrt{\varphi_1^2 + \varphi_2^2}$ is the radial mode as it feels the potential. The remaining angular component, $\vartheta$, is the direction perpendicular to the radial mode, corresponding to the degree of freedom on the vacuum manifold and thus the massless mode.
\footnote{$\theta_\pm$ and $\vartheta$ are identified with the phases of the following $SU(2)$ transformation: 
\begin{align}
    \phi = e^{i\theta_-\sigma^3} e^{-i\vartheta\sigma^2} e^{i\theta_+\sigma^3} \left(\begin{array}{c}
        \varphi_r \\
        0
    \end{array}\right) ,
\end{align}
where $\sigma^a$ $(a=1,2,3)$ are the Pauli matrices. 
}

The original canonical kinetic terms of the scalar sector can be expressed as
\begin{align}
\rho_K &= |\dot\Phi_1|^2 + |\dot\Phi_2|^2 \\[1mm] 
&=  \frac{1}{2} \dot\varphi_r^2 + \frac{1}{2} \varphi_r^2 (\dot\theta_+^2 + \dot\theta_-^2 + \dot\vartheta^2) + F(\varphi_r,\vartheta,\theta_+,\theta_-),
\end{align}
where 
\begin{align}
\label{F_specificform}
    F(\varphi_r,\vartheta,\theta_+,\theta_-) = \varphi_r^2 \cos(2\vartheta) \dot\theta_+ \dot\theta_-.
\end{align}
In practice, we first compute the time derivatives of $\varphi_c$ and $\theta_c$ as
\begin{align}
    \dot\varphi_c = \frac{2 {\rm Re}(\Phi_c^* \dot\Phi_c)}{\varphi_c} \,,~~
    \dot\theta_c = \frac{2 {\rm Im} (\Phi_c^* \dot\Phi_c)}{\varphi_c^2} \,.
\end{align}
Then, we obtain $\dot\varphi_r, \dot\vartheta$, and $\dot\theta_\pm$ as
\begin{align}
    &\dot\varphi_r = \frac{\varphi_1\dot\varphi_1 + \varphi_2\dot\varphi_2}{\varphi_r} \,, \\
    &\dot\vartheta = \frac{\varphi_1\dot\varphi_2 - \varphi_2\dot\varphi_1}{\varphi_r^2} \,, \\
    &\dot\theta_\pm = \frac{\dot\theta_1 \pm \dot\theta_2}{2} \,,
\end{align}
which allows us to compute the energy density and the number density of each mode\footnote{
Practically, we mask the region around the string core by multiplying each time-derivative term by $(\varphi_r/v)$ as follows,
\begin{align}
\dot\varphi_{r,{\rm mask}} = \frac{\varphi_r}{v}\dot\varphi_r,~~v \dot\vartheta_{\rm mask} = \varphi_r \dot\vartheta,~~v \dot\theta_{{\pm},{\rm mask}} = \varphi_r \dot\theta_\pm. 
\end{align}
Similarly, we also mask the point at which either $\varphi_1$ or $\varphi_2$ vanishes.
}.

Since the semilocal string can be regarded as a linear excitation of $\varphi_r$, massless modes can be emitted from the strings via the above interaction terms.
However, since the dynamics is fully nonlinear, the perturbative analysis cannot be applied in this case
\footnote{
The NG boson emission from the global string can be described by the effective Kalb-Ramond action. See the discussion in Sec.~\ref{sec:discussion}.
}.
Indeed, the massless NG modes can be excited by the fully nonlinear dynamics of semilocal strings extended over the horizon scale. Naively, the oscillation of horizon-sized loop can excite the NG modes with horizon-scale wavelength, but we need precise numerical simulations to follow such a nonlinear emission process.

\section{Numerical results} \label{sec:numerical}

We have performed numerical lattice simulations to solve the nonlinear system of Eqs.~(\ref{eq:EoM_Phi}) and (\ref{eq:EoM_A}) as in previous studies \cite{Achucarro:2005vpt,Achucarro:2013mga,Lopez-Eiguren:2017ucu}.
The lattice gauge formulation is adopted in which the discrete gauge symmetry is imposed on each lattice site \cite{Moriarty:1988fx}, and the Gauss's constraint (\ref{eq:Gauss_law}) is kept satisfied if it is initially satisfied.

Our simulations are performed with both ``the fat string regime'' and ``the physical string regime''. The physical string regime corresponds to the realistic case, but the resolution becomes worse with time due to the cosmic expansion. Thus, the simulation should be terminated when the lattice spacing becomes comparable to the string width or magnetic core size. In the fat string regime, on the other hand, the scalar self-coupling constant, $\lambda$, and the gauge coupling constant, $q$, are time-dependent so that the string width and the magnetic core size can grow with the cosmic expansion, and thus we can always resolve them.

As the setup of our simulations, the number of grid points and the initial boxsize are, respectively, $N^3 = 4096^3$ and $L = 0.5N v^{-1}$ ($96v^{-1}$) for the fat (physical) string regime.
The model parameters are set as $\lambda = 0.025$ and $q = 1$, which reads $\beta = 0.05$.
\footnote{To produce a sufficiently populated network, small values of $\beta$ are favored~\cite{Achucarro:2005vpt}. However, for very small $\beta$, the width of the magnetic flux and the scalar excitation become too different in scale. Therefore, in this work we adopt the range of beta used in Ref.~\cite{Achucarro:2013mga}. We show the results for another value of $\beta$ in App.~\ref{app:beta005}, which confirm that the qualitative behavior do not depends on $\beta$.}
The conformal time, $\tau$ defined by $dt = ad\tau$, is adopted as the time variable,
and we update the field variables using the second-order leap-frog method in our simulations. 
The stepsize is $0.2 \delta x$ with $\delta x = L/N$ being the lattice spacing. The initial conformal time is $v\tau_i = 1$ and the scale factor is normalized by $a(\tau_i) = 1$. Note that the scale factor is given by $a = v\tau$ in our setup (radiation-dominated universe).

The initial value of $\Phi$ at each lattice site is set with the Gaussian random field. 
To reduce the artificial initial gradient energy, we generate random numbers in the Fourier space with the ultraviolet (UV) cutoff, $k_{\rm UV} = 10 k_{\rm IR}$, where $k_{\rm IR} = 2 \pi / L$ is the infrared (IR) cutoff. We have confirmed that the result is not sensitive to the cutoff scale.
The initial values of the other fields are set as zero.

\begin{figure*}[tp]
\centering
\includegraphics [width = 16cm, clip]{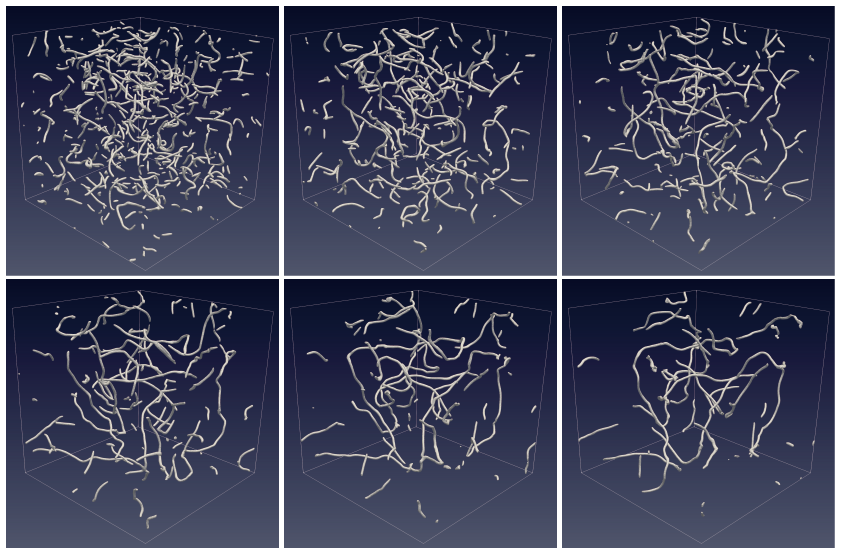}
\caption{
Snapshots of the semilocal string network for the fat string regime.
Time evolves from top left to top right and bottom left to bottom right, corresponding to the conformal time $\tau = 61v^{-1}$-$261v^{-1}$ with the interval $\Delta \tau = 40 v^{-1}$.
}
\label{fig:snapshot}
\end{figure*}

Identifying the semilocal string in the simulation is not trivial because one cannot use the winding number as an indicator of the string core, in contrast to the case with the Abelian-Higgs string. To find the semilocal string, we use the magnetic field concentration in the string \cite{Achucarro:2005vpt,Achucarro:2013mga}. To be more specific, we also perform the simulation with the Abelian-Higgs model to calculate the magnetic field strength averaged over all string segments, $B_{\rm core}^{({\rm AH})}$, whose positions can be identified by the local winding number. Then, the semilocal string can be detected at which the magnetic field strength exceeds the threshold value, $\gamma B_{\rm core}^{({\rm AH})}$ with $\gamma < 1$. We set $\gamma = 0.5$ in our numerical analysis.
Fig. \ref{fig:snapshot} shows snapshots of the evolution of the semilocal string network in our numerical simulations. One can see that strings have endpoints that correspond to global monopoles\cite{Hindmarsh:1992yy}.
Interaction between monopoles leads to the merging of segments, whereas small segments shrink and disappear due to the string tension. By the end of the simulation, almost all segments have vanished from the simulation box except for those larger than the horizon scale.

To see the scaling behavior of the semilocal string network, we calculate the total length of the string, $\ell_{\rm str}$ with the help of the Abelian-Higgs simulation as follows. 
The total magnetic field energy, $E_{B}$, can be obtained in both the semilocal and the Abelian-Higgs string cases. Assuming that the magnetic field is confined in the string core, it is given by the product $E_B^{\rm (X)} = \mu_B^{\rm (X)} \ell_{\rm str}^{\rm (X)}$, where $\mu_B$ is the magnetic energy per unit length and X is either SL (semilocal) or AH (Abelian-Higgs). 
For a straight static string, $g(r)$ in Eq.~\eqref{semilocal_string_solution} is identical to that of the Abelian-Higgs string when $\beta$ is the same~\cite{Vachaspati:1991dz}. Then we take $\mu_B^{\rm (SL)} = \mu_B^{\rm (AH)}$ and obtain $\ell_{\rm str}^{\rm (SL)} = (E_B^{\rm (SL)}/E_B^{\rm (AH)}) \ell_{\rm str}^{\rm (AH)}$.
Note that the length of the Abelian-Higgs string can be calculated by using the local winding number.

\begin{figure}[tp]
\centering
\includegraphics [width = 8.5cm, clip]{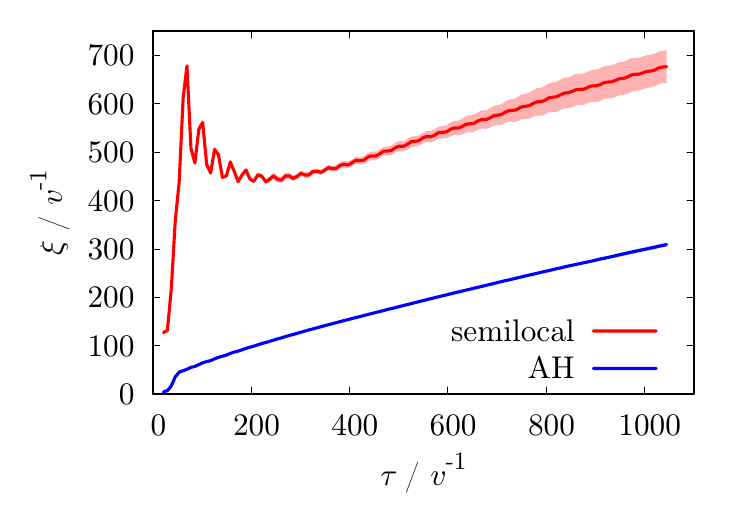}
\label{subfig:nk_NG1}
\includegraphics [width = 8.5cm, clip]{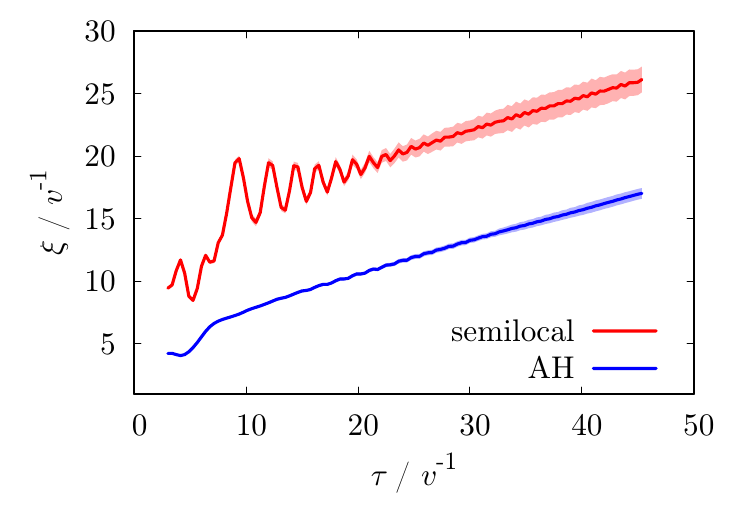}
\label{subfig:nk_NG2}
\caption{
Time evolution of the mean string separation for the semilocal string (red) and the Abelian-Higgs string (blue) in the fat/physical string regime (top/bottom panel). We have taken $\lambda = 0.025$ and $q = 1$ ($\beta = 0.05$). The solid curve and shaded region represent respectively the mean value and 1-$\sigma$ error with 10 independent simulations.
}
\label{fig:scaling}
\end{figure}

The scaling property can be characterized by the mean string separation defined by
\begin{align}
\xi_{\rm phys} = \sqrt{\frac{V}{\ell_{\rm str}}}
\end{align}
where $V (= a^3 L^3)$ is the physical volume of the simulation box. It increases linearly with time if the network is in the scaling regime.
Fig.~\ref{fig:scaling} shows the time evolution of the comoving mean string separation, $\xi = \xi_{\rm phys}/a$, of semilocal strings (red) and Abelian-Higgs strings (blue). 
The figure clearly demonstrates the scaling behavior of the semilocal string, confirming the results in the literature \cite{Achucarro:2005vpt,Achucarro:2013mga,Lopez-Eiguren:2017ucu}. 
The oscillatory behavior of $\xi$ at early $\tau$ in the semilocal string is a consequence of spurious transient oscillations of the background magnetic field induced by the initial configuration of $\phi$ since the simulation starts with the zero gauge fields. 
As $\tau$ increases, these artificial oscillations are damped, and we can observe the scaling behavior of the network at late times.

\begin{figure}[tp]
\centering
\includegraphics [width = 8.5cm, clip]{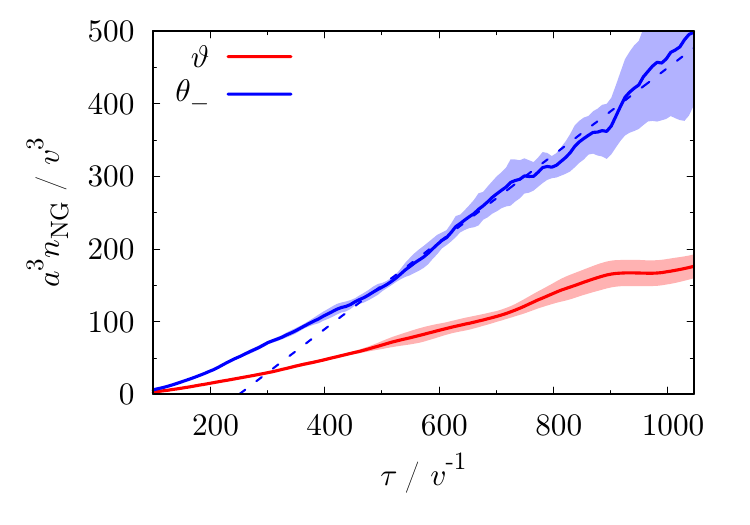}
\label{subfig:nk_NG1}
\includegraphics [width = 8.5cm, clip]{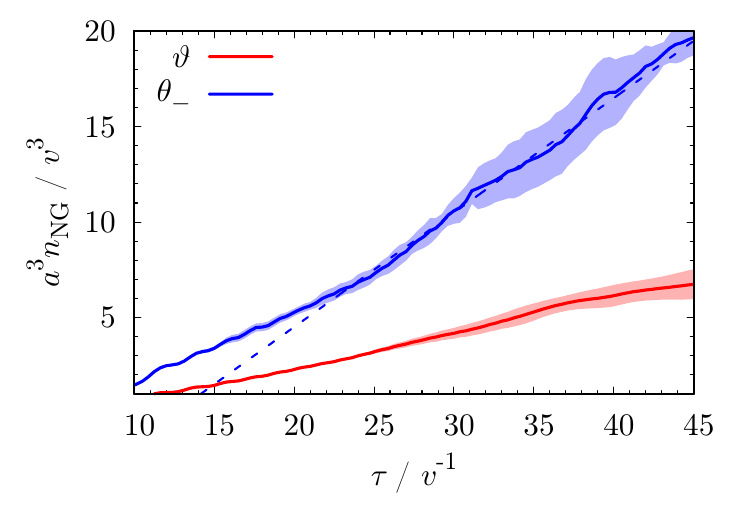}
\label{subfig:nk_NG2}
\caption{
Time evolution of the comoving number density of two NG modes for the fat/physical string regime in the top/bottom panel. The solid curve and shaded region represent respectively the mean value and 1-$\sigma$ error with 10 independent simulations.
The dashed line is the linear fitting line, $Av\tau + B$ with $(A,B) = (0.6,-150)$ (top), $(0.6,-7.5)$ (bottom). We have taken $\lambda = 0.025$ and $q = 1$ ($\beta = 0.05$).
}
\label{fig:evolve}
\end{figure}

Masless NG bosons can be emitted from the network of semilocal strings, mainly through the decay of loops. Initially, the loop has a horizon size, and its oscillations can excite the NG modes with Hubble-scale frequencies.
Fig. \ref{fig:evolve} shows the time evolution of the comoving number density of massless NG bosons, $a^3 n_{\rm NG}$, in the fat/physical string regime (top/bottom panel).
The figure demonstrates that the NG bosons are efficiently emitted from the network of semilocal strings. 
Its linear dependence on the conformal time supports the analytic estimation with the assumption that the emitted NG bosons has Hubble-scale frequencies, which is shown in the next section.

\begin{figure}[tp]
\centering
\includegraphics [width = 8.5cm, clip]{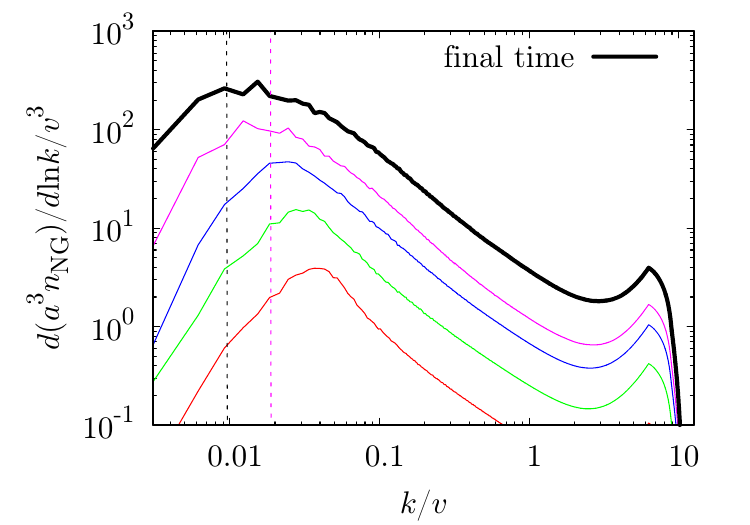}
\caption{
Evolution of the spectrum of NG modes in the fat string case for 
$v\tau = 85,\,149,\,277,\,533,\,1045$.
The $k$ on the horizontal axis denotes the comoving momentum. 
Time evolves from bottom to top and the thick black line corresponds to the final time. 
The dashed magenta and black lines correspond to $\bar{k}=10H$ at $v\tau = 533,\,1045$, respectively.
}
\label{fig:spectrum_evo}
\end{figure}

Fig.~\ref{fig:spectrum_evo} exhibits the evolution of the spectrum of the comoving number density of the NG mode (the sum of the $\vartheta$ and $\theta_-$ modes). It clearly shows that the low-energy peak grows and shifts to lower comoving momentum $k$ as time goes on.
In particular, the peak physical momentum is proportional to the Hubble parameter ($\bar{k}_{\rm peak} \sim 10 H$) when the network evolves in the scaling regime (see the dashed vertical lines). Note that the gradual shift of the peak wavenumber implies that the emission is not due to the decay of heavy modes.
We can also see a small peak in the high-$k$ region, which may be an artifact caused by the finite resolution.

\begin{figure}[tp]
\centering
\includegraphics [width = 8.5cm, clip]{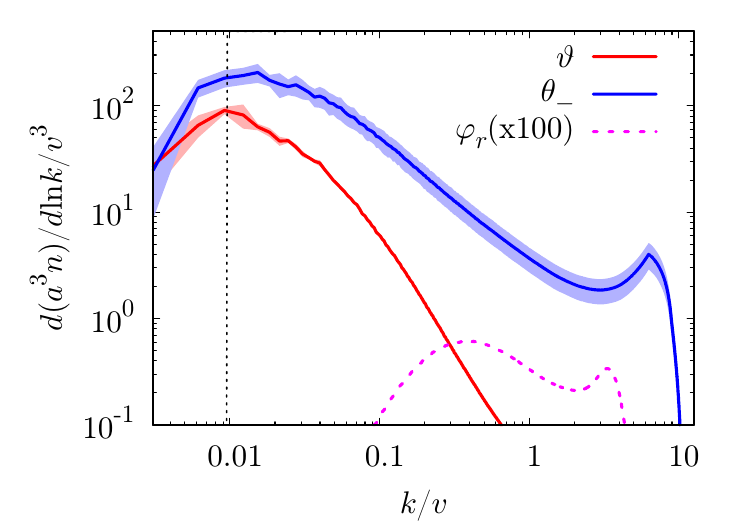}
\label{subfig:spectrum_fat}
\includegraphics [width = 8.5cm, clip]{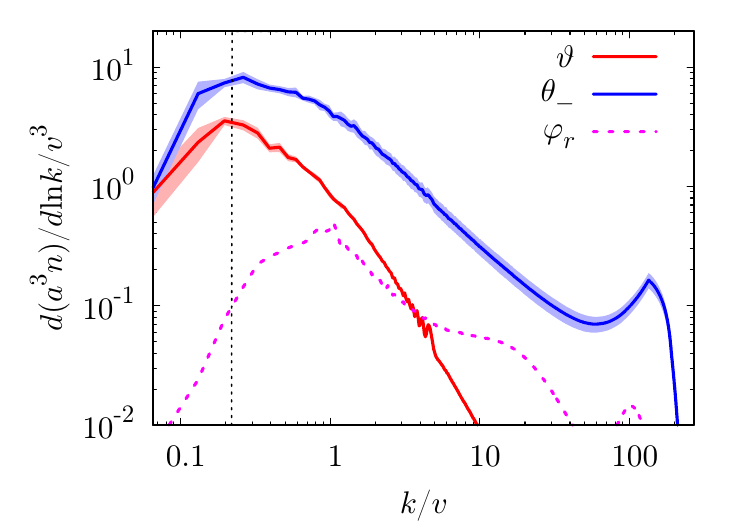}
\label{subfig:spectrum_phys}
\caption{
Spectra of the comoving number density for two NG modes, $\vartheta$ (solid red) and $\theta_-$ (solid blue), and the radial mode, $\varphi_r$ (dashed magenta), at the final time of the simulation in the fat/physical string regime (top/bottom panel). 
The solid curve and shaded region represent respectively the mean value and 1-$\sigma$ error with 10 independent simulations.
The dashed line corresponds to $\bar{k}=10H$ at that time.
}
\label{fig:spectra}
\end{figure}

Figs.~\ref{fig:spectra} show the spectra for individual massless NG modes, $\vartheta$, $\theta_-$, and the radial mode, $\varphi_r$, at the final time of the simulation. 
One can find that low-energy particles are dominantly produced, 
as in the case with the axion emission from global strings \cite{Saurabh:2020pqe,Baeza-Ballesteros:2023say} and the light dark photon emission from near-global Abelian-Higgs strings \cite{Kitajima:2022lre}. The peak wave number corresponds to the length scale $\alpha \ell_H$ with $\alpha$ and $\ell_H$ being, respectively, a numerical constant and the Hubble horizon scale. In both the fat and physical string regimes, we found $\alpha \sim 0.1$. 
Our results show that the production of the $\theta_-$-mode is more efficient than that of the $\vartheta$-mode. 
We attribute this difference to the asymmetric couplings of the massless NG modes to $\varphi_r$ as in Eq.~\eqref{F_specificform}. However, this asymmetry does not affect the qualitative nature of our conclusion.

\section{Pseudo-NG boson as dark matter} \label{sec:pNGBdm}
If NG bosons acquire the mass due to soft-breaking terms, they can take the role of cold dark matter.
Such a pseudo NG (pNG) dark matter model has been proposed in \cite{Abe:2022mlc,Abe:2024vxz}.
As in the case with the light dark photon emission \cite{Long:2019lwl,Kitajima:2022lre}, 
we assume that the pNG emission is efficient as long as its mass is smaller than the Hubble parameter but the emission stops when the Hubble parameter becomes comparable to the pNG boson mass.
At that time, the abundance of the pNG bosons is fixed.
The relic abundance of the pNG dark matter can be calculated from our numerical results as follows.

The typical momentum (energy) of the NG bosons emitted from the semilocal string network can be expressed as $\bar{k}/a = \bar{E}_{\rm NG} = (\alpha \ell_H)^{-1}$, where we implicitly assume that the particles are emitted by the decay of loops of typical size $\alpha \ell_H$. 
Then, the number density of the produced particles can be formally expressed as\cite{Long:2019lwl}
\begin{align}
n_{\rm NG} = \frac{8 \zeta \mu H}{\bar{E}_{\rm NG}/H},
\end{align}
where $\mu = \pi v^2$ is the string tension and $\zeta$ is the average number of long strings in a Hubble volume, defined by $\rho_{\rm str} = \zeta \mu/t^2$ with $\rho_{\rm str}$ being the energy density of long strings.\footnote{
In the case of the axion emission from global strings, the scaling violation, i.e. the logarithmic time dependence of the scaling density, has been reported by some groups \cite{Gorghetto:2018myk,Kawasaki:2018bzv,Buschmann:2021sdq,Saikawa:2024bta,Benabou:2024msj}, while Refs. \cite{Hindmarsh:2019csc,Hindmarsh:2021vih,Correia:2024cpk} show no scaling violation.
In the case of the semilocal string, the string profile is similar to that of the Abelian-Higgs string, and we do not consider the scaling violation.
}
Our numerical results in Fig. \ref{fig:evolve} show $a^3 n_{\rm NG} \propto a$, which reads $n_{\rm NG} \propto a^{-2} \propto H$. More specifically, we found $n_{\rm NG} \simeq 0.6 v^2 H$, assuming that $\theta_-$ serves as the dark matter. 
The relic abundance can be calculated by $n_{{\rm NG},0}/s_0 = n_{\rm NG} (t_*) /s(t_*)$ with $H(t_*) = m_{\rm NG}$, which leads to
\begin{align}
    \Omega_{\rm NG}h^2 &= \frac{m_{\rm NG} n_{{\rm NG},0}/s_0}{\rho_{{\rm cr},0}/s_0 h^{-2}} \nonumber \\
    &\simeq 0.2 \left( \frac{m_{\rm NG}}{10^{-13} {\rm eV}} \right)^{1/2} \left( \frac{v}{10^{14} {\rm GeV}} \right)^2.
\end{align}

\begin{figure}[tp]
\centering
\includegraphics [width = 8.5cm, clip]{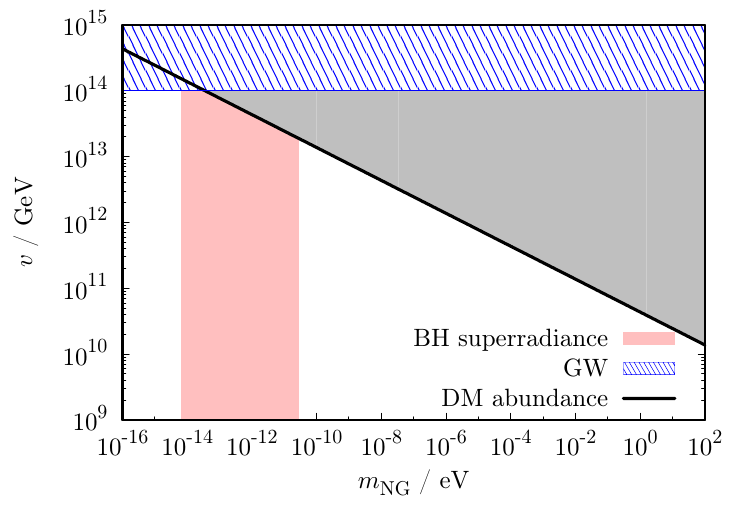}
\caption{
Allowed parameter region in $m_{\rm NG}$-$v$ plane. The current relic abundance of dark matter can be explained on the thick black line, i.e., $\Omega_{\rm NG} = \Omega_{\rm DM}$.
The gray shaded region is ruled out due to the overabundance of dark matter.
The red shaded and the blue hatched regions represent respectively the constraints from the black hole superradiance and the gravitational wave observations.
}
\label{fig:constraint}
\end{figure}

Fig.\ref{fig:constraint} shows the viable parameter region when the pNG boson accounts for the dominant or subdominant component of the dark matter. Note that we have imposed $v < 10^{14}\, {\rm GeV}$ from the observation of gravitational waves, taking into account the suppression of the gravitational wave emissions due to light particle emissions \cite{Kitajima:2022lre}. 
This is a rough estimation and detailed study for the gravitational wave emission from the semilocal string network is required, which is left for future work. 
From Fig.\ref{fig:constraint}, the pNG dark matter should be very light, with a lower bound $m_{\rm NG} \gtrsim 10^{-10}\, {\rm eV}$, if the pNG boson is the dominant dark matter. It should be emphasized that the scenario with the formation of the semilocal string network provides an alternative production mechanism of the light pNG dark matter.

\section{Discussion} \label{sec:discussion}

We have numerically shown that massless NG modes are efficiently produced from the semilocal string network. First, we have confirmed the scaling behavior of the network. Then, we have revealed that the comoving number density of emitted NG bosons grows proportional to the Hubble parameter, which is consistent with the analytic estimation. The spectrum of the comoving number density of NG bosons has a peak corresponding to the length scale of $0.1 \ell_H$ (horizon scale), which implies that a typical momentum of the emitted NG bosons is much smaller than the symmetry breaking scale.
If NG bosons acquire the mass due to soft-breaking terms, they can contribute to the dark matter component in the current universe. The mass of such a pNG dark matter can be as small as $10^{-10}\, {\rm eV}$.

Unlike the Abelian-Higgs string model, the gravitational wave emission from the network of semilocal strings may be suppressed due to the persistent emissions of massless NG bosons as in the case with the global string \cite{Chang:2019mza,Figueroa:2020lvo,Chang:2021afa,Baeza-Ballesteros:2023say}. However, the NG boson emission may become inefficient at some time if the NG boson has mass like the scenario in \cite{Long:2019lwl,Kitajima:2022lre,Kitajima:2023vre}. In this case, the string loses its energy only through the gravitational wave emissions in the late universe. Thus, the resultant gravitational wave spectrum may be modified significantly from the nearly scale-invariant one. We will analyze it in the subsequent paper.

Our simulation results show that the characteristic momentum of the emitted NG bosons is of order the Hubble scale. This behavior is analogous to the axion radiation from a global string, which is successfully described by a Kalb–Ramond field coupled to a Nambu–Goto string~\cite{Kalb:1974yc,Vilenkin:1986ku}. Motivated by this analogy, we expect that the NG-boson emission from a semilocal string can likewise be captured by antisymmetric tensor fields dual to $\theta_-$ and $\vartheta$. However, it is not obvious that these tensor fields couple to the string in the same manner as in the Kalb–Ramond action, since $\theta_-$ and $\vartheta$ do not wind around the string core. We also note that the doublet $n$ in Eq.~\eqref{semilocal_string_solution} can undergo dynamical fluctuations near the string core, which may provide an alternative mechanism for coupling the NG modes to the string. Clarifying the structure of the corresponding low-energy theory remains an open question.

If the SU(2) symmetry is the gauge symmetry rather than the global symmetry, the model predicts the Z-string\cite{Vachaspati:1992fi}. 
In this case, the NG modes we considered so far turn into the longitudinal modes of the massive gauge bosons, while a massless $U(1)$ gauge boson remains, corresponding to the residual U(1) gauge symmetry. This system also admits the energy loss due to the massless particle emissions and thus predicts light dark photon dark matter if the U(1) gauge symmetry is eventually broken, and characteristic signature in the gravitational wave spectrum.
The study on this model is left for future work.

\section*{Acknowledgments}
This work is supported by JSPS KAKENHI Grant No. JP24KJ1257 (Y.K.).
This work used computational resources of Fugaku supercomputer, provided by RIKEN Center for Computational Sciences, through the HPCI System Research Project (Project ID: hp240131, hp250177). 
The authors thank Tomohiro Abe for useful discussions about the pNG dark matter.

\appendix

\section{Case with different $\beta$} \label{app:beta005}

\begin{figure}[tp]
\centering
\includegraphics [width = 8.5cm, clip]{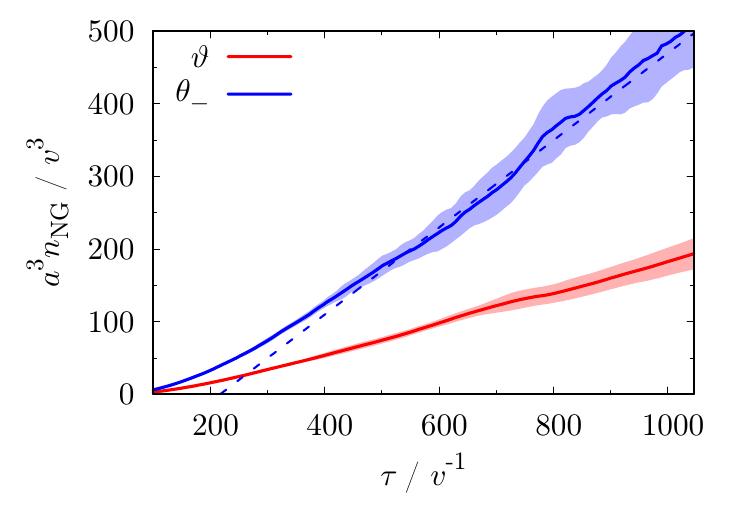}
\label{subfig:evolve_n_fat2}
\includegraphics [width = 8.5cm, clip]{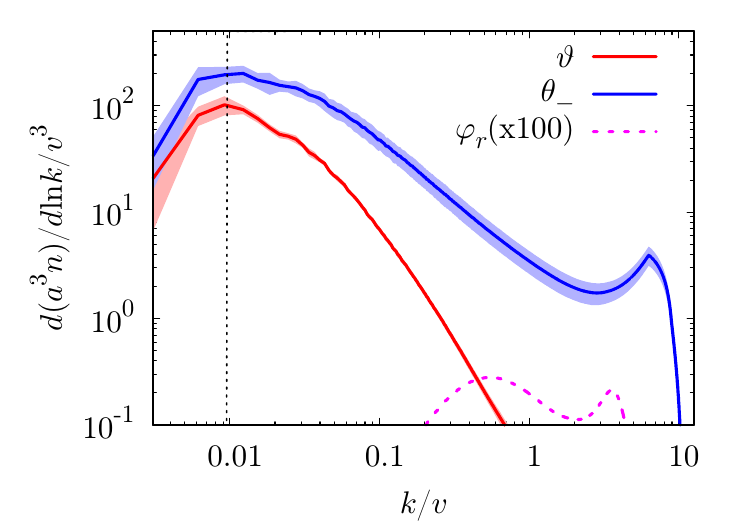}
\label{subfig:spectrum_fat2}
\caption{
The evolution of the comoving number density of two NG modes (top panel) and their spectra at the final time of the simulation (bottom panel) for the fat string case with $\lambda = 0.05$ and $q=1$ ($\beta = 0.1$). The graphical elements (colored regions, dashed lines, etc.) follow the same convention as in Figs.~\ref{fig:evolve} and \ref{fig:spectra}. 
}
\label{fig:beta005}
\end{figure}

Here we show the results of the fat string simulation in the case with a different value of $\beta$. 
The model parameters are set as $\lambda = 0.05$ and $q = 1$, which reads $\beta = 0.1$. 
Fig.~\ref{fig:beta005} shows the time evolution of the comoving number density of NG bosons and their spectra at the final time of the simulation. 
We find that $n_{\rm NG}a^3$ grows linearly with $\tau$, in the same manner as in Fig.~\ref{fig:evolve}. 
The typucal momenta of the emitted NG bosons are also around $10H$. 
These results indicate that our conclusions do not depend on the value of $\beta$.


\bibliography{semilocal.bib}

@book{Vilenkin:2000jqa,
	author = {Vilenkin, A. and Shellard, E. P. S.},
	isbn = {978-0-521-65476-0},
	month = {7},
	publisher = {Cambridge University Press},
	title = {{Cosmic Strings and Other Topological Defects}},
	year = {2000}}

@article{Buschmann:2021sdq,
	archiveprefix = {arXiv},
	author = {Buschmann, Malte and Foster, Joshua W. and Hook, Anson and Peterson, Adam and Willcox, Don E. and Zhang, Weiqun and Safdi, Benjamin R.},
	doi = {10.1038/s41467-022-28669-y},
	eprint = {2108.05368},
	journal = {Nature Commun.},
	number = {1},
	pages = {1049},
	primaryclass = {hep-ph},
	title = {{Dark matter from axion strings with adaptive mesh refinement}},
	volume = {13},
	year = {2022}}

@article{Kawasaki:2018bzv,
	archiveprefix = {arXiv},
	author = {Kawasaki, Masahiro and Sekiguchi, Toyokazu and Yamaguchi, Masahide and Yokoyama, Jun'ichi},
	doi = {10.1093/ptep/pty098},
	eprint = {1806.05566},
	journal = {PTEP},
	number = {9},
	pages = {091E01},
	primaryclass = {hep-ph},
	reportnumber = {RESCEU-8/18, RESCEU-8-18},
	title = {{Long-term dynamics of cosmological axion strings}},
	volume = {2018},
	year = {2018}}

@article{Saikawa:2024bta,
	archiveprefix = {arXiv},
	author = {Saikawa, Ken'ichi and Redondo, Javier and Vaquero, Alejandro and Kaltschmidt, Mathieu},
	doi = {10.1088/1475-7516/2024/10/043},
	eprint = {2401.17253},
	journal = {JCAP},
	pages = {043},
	primaryclass = {hep-ph},
	reportnumber = {KANAZAWA-24-02, MPP-2024-18},
	title = {{Spectrum of global string networks and the axion dark matter mass}},
	volume = {10},
	year = {2024}}

@article{Gorghetto:2020qws,
	archiveprefix = {arXiv},
	author = {Gorghetto, Marco and Hardy, Edward and Villadoro, Giovanni},
	doi = {10.21468/SciPostPhys.10.2.050},
	eprint = {2007.04990},
	journal = {SciPost Phys.},
	number = {2},
	pages = {050},
	primaryclass = {hep-ph},
	title = {{More axions from strings}},
	volume = {10},
	year = {2021}}

@article{Gorghetto:2021fsn,
	archiveprefix = {arXiv},
	author = {Gorghetto, Marco and Hardy, Edward and Nicolaescu, Horia},
	doi = {10.1088/1475-7516/2021/06/034},
	eprint = {2101.11007},
	journal = {JCAP},
	pages = {034},
	primaryclass = {hep-ph},
	title = {{Observing invisible axions with gravitational waves}},
	volume = {06},
	year = {2021}}

@article{Abe:2024vxz,
	archiveprefix = {arXiv},
	author = {Abe, Tomohiro and Hamada, Yu and Tsumura, Koji},
	doi = {10.1007/JHEP05(2024)076},
	eprint = {2401.02397},
	journal = {JHEP},
	pages = {076},
	primaryclass = {hep-ph},
	reportnumber = {KEK-TH-2589, DESY-24-002, KYUSHU-HET-277},
	title = {{A model of pseudo-Nambu-Goldstone dark matter with two complex scalars}},
	volume = {05},
	year = {2024}}

@article{Preskill:1992bf,
	archiveprefix = {arXiv},
	author = {Preskill, John},
	doi = {10.1103/PhysRevD.46.4218},
	eprint = {hep-ph/9206216},
	journal = {Phys. Rev. D},
	pages = {4218--4231},
	reportnumber = {CALT-68-1787},
	title = {{Semilocal defects}},
	volume = {46},
	year = {1992}}

@article{Vilenkin:1986ku,
	author = {Vilenkin, Alexander and Vachaspati, Tanmay},
	doi = {10.1103/PhysRevD.35.1138},
	journal = {Phys. Rev. D},
	pages = {1138},
	reportnumber = {TUTP-86-16},
	title = {{Radiation of Goldstone Bosons From Cosmic Strings}},
	volume = {35},
	year = {1987}}

@article{Chang:2019mza,
	archiveprefix = {arXiv},
	author = {Chang, Chia-Feng and Cui, Yanou},
	doi = {10.1016/j.dark.2020.100604},
	eprint = {1910.04781},
	journal = {Phys. Dark Univ.},
	pages = {100604},
	primaryclass = {hep-ph},
	title = {{Stochastic Gravitational Wave Background from Global Cosmic Strings}},
	volume = {29},
	year = {2020}}

@article{Figueroa:2020lvo,
	archiveprefix = {arXiv},
	author = {Figueroa, Daniel G. and Hindmarsh, Mark and Lizarraga, Joanes and Urrestilla, Jon},
	doi = {10.1103/PhysRevD.102.103516},
	eprint = {2007.03337},
	journal = {Phys. Rev. D},
	number = {10},
	pages = {103516},
	primaryclass = {astro-ph.CO},
	title = {{Irreducible background of gravitational waves from a cosmic defect network: update and comparison of numerical techniques}},
	volume = {102},
	year = {2020}}

@article{Baeza-Ballesteros:2023say,
	archiveprefix = {arXiv},
	author = {Baeza-Ballesteros, Jorge and Copeland, Edmund J. and Figueroa, Daniel G. and Lizarraga, Joanes},
	doi = {10.1103/PhysRevD.110.043522},
	eprint = {2308.08456},
	journal = {Phys. Rev. D},
	number = {4},
	pages = {043522},
	primaryclass = {astro-ph.CO},
	title = {{Gravitational wave emission from a cosmic string loop: Global case}},
	volume = {110},
	year = {2024}}

@article{Chang:2021afa,
	archiveprefix = {arXiv},
	author = {Chang, Chia-Feng and Cui, Yanou},
	doi = {10.1007/JHEP03(2022)114},
	eprint = {2106.09746},
	journal = {JHEP},
	pages = {114},
	primaryclass = {hep-ph},
	title = {{Gravitational waves from global cosmic strings and cosmic archaeology}},
	volume = {03},
	year = {2022}}

@article{Saurabh:2020pqe,
	archiveprefix = {arXiv},
	author = {Saurabh, Ayush and Vachaspati, Tanmay and Pogosian, Levon},
	doi = {10.1103/PhysRevD.101.083522},
	eprint = {2001.01030},
	journal = {Phys. Rev. D},
	number = {8},
	pages = {083522},
	primaryclass = {hep-ph},
	title = {{Decay of Cosmic Global String Loops}},
	volume = {101},
	year = {2020}}

@article{Kitajima:2023vre,
	archiveprefix = {arXiv},
	author = {Kitajima, Naoya and Nakayama, Kazunori},
	doi = {10.1016/j.physletb.2023.138213},
	eprint = {2306.17390},
	journal = {Phys. Lett. B},
	pages = {138213},
	primaryclass = {hep-ph},
	reportnumber = {TU-1199, KEK-QUP-2023-0015},
	title = {{Nanohertz gravitational waves from cosmic strings and dark photon dark matter}},
	volume = {846},
	year = {2023}}

@article{Kitajima:2022lre,
	archiveprefix = {arXiv},
	author = {Kitajima, Naoya and Nakayama, Kazunori},
	doi = {10.1007/JHEP08(2023)068},
	eprint = {2212.13573},
	journal = {JHEP},
	pages = {068},
	primaryclass = {hep-ph},
	reportnumber = {TU-1175, KEK-QUP-2022-0021},
	title = {{Dark photon dark matter from cosmic strings and gravitational wave background}},
	volume = {08},
	year = {2023}}

@article{Long:2019lwl,
	archiveprefix = {arXiv},
	author = {Long, Andrew J. and Wang, Lian-Tao},
	doi = {10.1103/PhysRevD.99.063529},
	eprint = {1901.03312},
	journal = {Phys. Rev. D},
	number = {6},
	pages = {063529},
	primaryclass = {hep-ph},
	title = {{Dark Photon Dark Matter from a Network of Cosmic Strings}},
	volume = {99},
	year = {2019}}

@article{Lopez-Eiguren:2017ucu,
	archiveprefix = {arXiv},
	author = {Lopez-Eiguren, A. and Urrestilla, J. and Ach\'ucarro, A. and Avgoustidis, A. and Martins, C. J. A. P.},
	doi = {10.1103/PhysRevD.96.023526},
	eprint = {1704.00991},
	journal = {Phys. Rev. D},
	number = {2},
	pages = {023526},
	primaryclass = {hep-ph},
	title = {{Evolution of Semilocal String Networks: II. Velocity estimators}},
	volume = {96},
	year = {2017}}

@article{Achucarro:2005vpt,
	archiveprefix = {arXiv},
	author = {Achucarro, Ana and Salmi, Petja and Urrestilla, Jon},
	doi = {10.1103/PhysRevD.75.121703},
	eprint = {astro-ph/0512487},
	journal = {Phys. Rev. D},
	pages = {121703},
	title = {{Semilocal cosmic string networks}},
	volume = {75},
	year = {2007}}

@article{Achucarro:2013mga,
	archiveprefix = {arXiv},
	author = {Ach\'ucarro, A. and Avgoustidis, A. and Leite, A. M. M. and Lopez-Eiguren, A. and Martins, C. J. A. P. and Nunes, A. S. and Urrestilla, J.},
	doi = {10.1103/PhysRevD.89.063503},
	eprint = {1312.2123},
	journal = {Phys. Rev. D},
	number = {6},
	pages = {063503},
	primaryclass = {hep-ph},
	title = {{Evolution of semilocal string networks: Large-scale properties}},
	volume = {89},
	year = {2014}}

@article{Achucarro:1998ux,
	archiveprefix = {arXiv},
	author = {Achucarro, Ana and Borrill, Julian and Liddle, Andrew R},
	doi = {10.1103/PhysRevLett.82.3742},
	eprint = {hep-ph/9802306},
	journal = {Phys. Rev. Lett.},
	pages = {3742--3745},
	reportnumber = {EHU-FT-9801, UG-3-98, CFPA-98-TH-02, SUSSEX-AST-98-2-3},
	title = {{The Formation rate of semilocal strings}},
	volume = {82},
	year = {1999}}

@article{Achucarro:1999it,
	archiveprefix = {arXiv},
	author = {Achucarro, Ana and Vachaspati, Tanmay},
	doi = {10.1016/S0370-1573(99)00103-9},
	eprint = {hep-ph/9904229},
	journal = {Phys. Rept.},
	pages = {347--426},
	reportnumber = {EHU-FT-9808A, CWRU-P34-1998},
	title = {{Semilocal and electroweak strings}},
	volume = {327},
	year = {2000}}

@article{Vachaspati:1991dz,
	author = {Vachaspati, T. and Achucarro, A.},
	doi = {10.1103/PhysRevD.44.3067},
	journal = {Phys. Rev. D},
	pages = {3067--3071},
	title = {{Semilocal cosmic strings}},
	volume = {44},
	year = {1991}}

@article{Gibbons:1992gt,
	archiveprefix = {arXiv},
	author = {Gibbons, G. W. and Ortiz, M. E. and Ruiz Ruiz, F. and Samols, T. M.},
	doi = {10.1016/0550-3213(92)90097-U},
	eprint = {hep-th/9203023},
	journal = {Nucl. Phys. B},
	pages = {127--144},
	reportnumber = {MIT-CTP-2063, DAMTP-R-92-7, NBI-HE-92-XX, DAMTP-HEP-92-09},
	title = {{Semilocal strings and monopoles}},
	volume = {385},
	year = {1992}}

@article{Hindmarsh:1991jq,
	author = {Hindmarsh, Mark},
	doi = {10.1103/PhysRevLett.68.1263},
	journal = {Phys. Rev. Lett.},
	pages = {1263--1266},
	reportnumber = {NCL-91-TP7},
	title = {{Existence and stability of semilocal strings}},
	volume = {68},
	year = {1992}}

@article{Hindmarsh:1992yy,
	archiveprefix = {arXiv},
	author = {Hindmarsh, Mark},
	doi = {10.1016/0550-3213(93)90681-E},
	eprint = {hep-ph/9206229},
	journal = {Nucl. Phys. B},
	pages = {461--492},
	reportnumber = {DAMTP-HEP-92-24, NSF-ITP-92-75},
	title = {{Semilocal topological defects}},
	volume = {392},
	year = {1993}}

@article{Vachaspati:1992fi,
    author = "Vachaspati, Tanmay",
    title = "{Vortex solutions in the Weinberg-Salam model}",
    reportNumber = "TUTP-91-12",
    doi = "10.1103/PhysRevLett.68.1977",
    journal = "Phys. Rev. Lett.",
    volume = "68",
    pages = "1977--1980",
    year = "1992",
    note = "[Erratum: Phys.Rev.Lett. 69, 216 (1992)]"
}

@article{Abe:2022mlc,
    author = "Abe, Tomohiro and Hamada, Yu",
    title = "{A model of pseudo-Nambu{\textendash}Goldstone dark matter from a softly broken SU(2) global symmetry with a U(1) gauge symmetry}",
    eprint = "2205.11919",
    archivePrefix = "arXiv",
    primaryClass = "hep-ph",
    reportNumber = "KEK-TH-2428",
    doi = "10.1093/ptep/ptad021",
    journal = "PTEP",
    volume = "2023",
    number = "3",
    pages = "033B04",
    year = "2023"
}

@article{Vilenkin:1981bx,
	author = {Vilenkin, A.},
	doi = {10.1016/0370-2693(81)91144-8},
	journal = {Phys. Lett. B},
	pages = {47--50},
	title = {{Gravitational radiation from cosmic strings}},
	volume = {107},
	year = {1981}}

@article{Vachaspati:1984gt,
	author = {Vachaspati, Tanmay and Vilenkin, Alexander},
	doi = {10.1103/PhysRevD.31.3052},
	journal = {Phys. Rev. D},
	pages = {3052},
	reportnumber = {HUTP-84/A065},
	title = {{Gravitational Radiation from Cosmic Strings}},
	volume = {31},
	year = {1985}}

@article{Kibble:1984hp,
	author = {Kibble, T. W. B.},
	doi = {10.1016/0550-3213(85)90596-6},
	editor = {Baier, R. and Satz, H.},
	journal = {Nucl. Phys. B},
	note = {[Erratum: Nucl.Phys.B 261, 750 (1985)]},
	pages = {227},
	reportnumber = {IMPERIAL-TP-83-84-54},
	title = {{Evolution of a system of cosmic strings}},
	volume = {252},
	year = {1985}}

@article{Bennett:1985qt,
	author = {Bennett, David P.},
	doi = {10.1103/PhysRevD.33.872},
	journal = {Phys. Rev. D},
	note = {[Erratum: Phys.Rev.D 34, 3932 (1986)]},
	pages = {872},
	reportnumber = {SLAC-PUB-3743},
	title = {{The evolution of cosmic strings}},
	volume = {33},
	year = {1986}}

@article{Bennett:1986zn,
	author = {Bennett, David P.},
	doi = {10.1103/PhysRevD.34.3592},
	journal = {Phys. Rev. D},
	pages = {3592},
	reportnumber = {SLAC-PUB-3989},
	title = {{Evolution of cosmic strings. 2.}},
	volume = {34},
	year = {1986}}

@article{Bennett:1989yp,
	author = {Bennett, David P. and Bouchet, Francois R.},
	doi = {10.1103/PhysRevD.41.2408},
	journal = {Phys. Rev. D},
	pages = {2408},
	reportnumber = {PUPT-89-1137, UCRL-102446},
	title = {{High resolution simulations of cosmic string evolution. 1. Network evolution}},
	volume = {41},
	year = {1990}}

@article{Cui:2017ufi,
	archiveprefix = {arXiv},
	author = {Cui, Yanou and Lewicki, Marek and Morrissey, David E. and Wells, James D.},
	doi = {10.1103/PhysRevD.97.123505},
	eprint = {1711.03104},
	journal = {Phys. Rev. D},
	number = {12},
	pages = {123505},
	primaryclass = {hep-ph},
	reportnumber = {KCL-PH-TH-2017-51},
	title = {{Cosmic Archaeology with Gravitational Waves from Cosmic Strings}},
	volume = {97},
	year = {2018}}

@article{Caprini:2018mtu,
	archiveprefix = {arXiv},
	author = {Caprini, Chiara and Figueroa, Daniel G.},
	doi = {10.1088/1361-6382/aac608},
	eprint = {1801.04268},
	journal = {Class. Quant. Grav.},
	number = {16},
	pages = {163001},
	primaryclass = {astro-ph.CO},
	title = {{Cosmological Backgrounds of Gravitational Waves}},
	volume = {35},
	year = {2018}}

@article{Auclair:2019wcv,
	archiveprefix = {arXiv},
	author = {Auclair, Pierre and others},
	doi = {10.1088/1475-7516/2020/04/034},
	eprint = {1909.00819},
	journal = {JCAP},
	pages = {034},
	primaryclass = {astro-ph.CO},
	title = {{Probing the gravitational wave background from cosmic strings with LISA}},
	volume = {04},
	year = {2020}}

@article{Kibble:1976sj,
	author = {Kibble, T. W. B.},
	doi = {10.1088/0305-4470/9/8/029},
	journal = {J. Phys. A},
	pages = {1387--1398},
	reportnumber = {ICTP/75/5},
	title = {{Topology of Cosmic Domains and Strings}},
	volume = {9},
	year = {1976}}

@article{Davis:1986xc,
	author = {Davis, Richard Lynn},
	doi = {10.1016/0370-2693(86)90300-X},
	journal = {Phys. Lett. B},
	pages = {225--230},
	reportnumber = {SLAC-PUB-3895},
	title = {{Cosmic Axions from Cosmic Strings}},
	volume = {180},
	year = {1986}}

@article{Garfinkle:1987yw,
	author = {Garfinkle, David and Vachaspati, Tanmay},
	doi = {10.1103/PhysRevD.36.2229},
	journal = {Phys. Rev. D},
	pages = {2229},
	reportnumber = {Print-87-0513 (WASH.U.,ST.LOUIS)},
	title = {{Radiation From Kinky, Cuspless Cosmic Loops}},
	volume = {36},
	year = {1987}}

@article{Yamaguchi:1998gx,
	archiveprefix = {arXiv},
	author = {Yamaguchi, Masahide and Kawasaki, M. and Yokoyama, Jun'ichi},
	doi = {10.1103/PhysRevLett.82.4578},
	eprint = {hep-ph/9811311},
	journal = {Phys. Rev. Lett.},
	pages = {4578--4581},
	reportnumber = {UTAP-313, YITP-98-73},
	title = {{Evolution of axionic strings and spectrum of axions radiated from them}},
	volume = {82},
	year = {1999}}

@article{Hagmann:2000ja,
	archiveprefix = {arXiv},
	author = {Hagmann, C. and Chang, Sanghyeon and Sikivie, P.},
	doi = {10.1103/PhysRevD.63.125018},
	eprint = {hep-ph/0012361},
	journal = {Phys. Rev. D},
	pages = {125018},
	reportnumber = {UFIFT-HEP-00-33},
	title = {{Axion radiation from strings}},
	volume = {63},
	year = {2001}}

@article{Hiramatsu:2010yu,
	archiveprefix = {arXiv},
	author = {Hiramatsu, Takashi and Kawasaki, Masahiro and Sekiguchi, Toyokazu and Yamaguchi, Masahide and Yokoyama, Jun'ichi},
	doi = {10.1103/PhysRevD.83.123531},
	eprint = {1012.5502},
	journal = {Phys. Rev. D},
	pages = {123531},
	primaryclass = {hep-ph},
	reportnumber = {IPMU10-0229, RESCEU-29-10, YITP-10-111},
	title = {{Improved estimation of radiated axions from cosmological axionic strings}},
	volume = {83},
	year = {2011}}

@article{Fleury:2015aca,
	archiveprefix = {arXiv},
	author = {Fleury, Leesa and Moore, Guy D.},
	doi = {10.1088/1475-7516/2016/01/004},
	eprint = {1509.00026},
	journal = {JCAP},
	pages = {004},
	primaryclass = {hep-ph},
	title = {{Axion dark matter: strings and their cores}},
	volume = {01},
	year = {2016}}

@article{Gorghetto:2018myk,
	archiveprefix = {arXiv},
	author = {Gorghetto, Marco and Hardy, Edward and Villadoro, Giovanni},
	doi = {10.1007/JHEP07(2018)151},
	eprint = {1806.04677},
	journal = {JHEP},
	pages = {151},
	primaryclass = {hep-ph},
	title = {{Axions from Strings: the Attractive Solution}},
	volume = {07},
	year = {2018}}

@article{Kim:2024wku,
	archiveprefix = {arXiv},
	author = {Kim, Heejoo and Park, Junghyeon and Son, Minho},
	doi = {10.1007/JHEP07(2024)150},
	eprint = {2402.00741},
	journal = {JHEP},
	pages = {150},
	primaryclass = {hep-ph},
	title = {{Axion dark matter from cosmic string network}},
	volume = {07},
	year = {2024}}

@article{Kim:2024dtq,
	archiveprefix = {arXiv},
	author = {Kim, Heejoo and Son, Minho},
	doi = {10.1007/JHEP07(2025)052},
	eprint = {2411.08455},
	journal = {JHEP},
	pages = {052},
	primaryclass = {hep-ph},
	title = {{More scalings from cosmic strings}},
	volume = {07},
	year = {2025}}

@article{Benabou:2024msj,
	archiveprefix = {arXiv},
	author = {Benabou, Joshua N. and Buschmann, Malte and Foster, Joshua W. and Safdi, Benjamin R.},
	doi = {10.1103/6v21-d6sj},
	eprint = {2412.08699},
	journal = {Phys. Rev. Lett.},
	number = {24},
	pages = {241003},
	primaryclass = {hep-ph},
	reportnumber = {FERMILAB-PUB-24-0912-T},
	title = {{Axion Mass Prediction from Adaptive Mesh Refinement Cosmological Lattice Simulations}},
	volume = {134},
	year = {2025}}

@article{Moriarty:1988fx,
	author = {Moriarty, K. J. M. and Myers, Eric and Rebbi, Claudio},
	doi = {10.1016/0370-2693(88)90674-0},
	journal = {Phys. Lett. B},
	pages = {411--418},
	reportnumber = {BUHEP-88-6},
	title = {{Dynamical Interactions of Flux Vortices in Superconductors}},
	volume = {207},
	year = {1988}}

@article{Hindmarsh:2019csc,
	archiveprefix = {arXiv},
	author = {Hindmarsh, Mark and Lizarraga, Joanes and Lopez-Eiguren, Asier and Urrestilla, Jon},
	doi = {10.1103/PhysRevLett.124.021301},
	eprint = {1908.03522},
	journal = {Phys. Rev. Lett.},
	number = {2},
	pages = {021301},
	primaryclass = {astro-ph.CO},
	reportnumber = {HIP-2020-1/TH},
	title = {{Scaling Density of Axion Strings}},
	volume = {124},
	year = {2020}}

@article{Hindmarsh:2021vih,
	archiveprefix = {arXiv},
	author = {Hindmarsh, Mark and Lizarraga, Joanes and Lopez-Eiguren, Asier and Urrestilla, Jon},
	doi = {10.1103/PhysRevD.103.103534},
	eprint = {2102.07723},
	journal = {Phys. Rev. D},
	number = {10},
	pages = {103534},
	primaryclass = {astro-ph.CO},
	reportnumber = {HIP-2021-7/TH},
	title = {{Approach to scaling in axion string networks}},
	volume = {103},
	year = {2021}}

@article{Correia:2024cpk,
	archiveprefix = {arXiv},
	author = {Correia, Jos{\'e} and Hindmarsh, Mark and Lizarraga, Joanes and Lopez-Eiguren, Asier and Rummukainen, Kari and Urrestilla, Jon},
	doi = {10.1103/PhysRevD.111.063532},
	eprint = {2410.18064},
	journal = {Phys. Rev. D},
	number = {6},
	pages = {063532},
	primaryclass = {hep-ph},
	title = {{Scaling density of axion strings in terasite simulations}},
	volume = {111},
	year = {2025}}

@article{Kitajima:2025jct,
    author = "Kitajima, Naoya and Uwabo-Niibo, Michiru",
    title = "{Multimodal axion emissions from Abelian-Higgs cosmic strings}",
    eprint = "2510.10708",
    archivePrefix = "arXiv",
    primaryClass = "hep-ph",
    month = "10",
    year = "2025"
}

@article{Kalb:1974yc,
    author = "Kalb, Michael and Ramond, Pierre",
    title = "{Classical direct interstring action}",
    doi = "10.1103/PhysRevD.9.2273",
    journal = "Phys. Rev. D",
    volume = "9",
    pages = "2273--2284",
    year = "1974"
}

\end{document}